# Solar Polar Diamond Explorer (SPDEx): Understanding the Origins of Solar Activity Using a New Perspective

A. Vourlidas[1], P. C. Liewer[2], M. Velli[3], D. Webb[4]

[1]*Johns Hopkins University Applied Physics Laboratory, Laurel, MD, USA*
[2]*Jet Propulsion Laboratory, California Institute of Technology, Pasadena, CA*
[3]*University of California Los Angeles, Los Angeles, CA*
[4]*Institute of Scientific Research, Boston College, Chestnut Hill, MA*

**Abstract**

Our knowledge of the Sun, its atmosphere, long term activity and space weather potential is severely limited by the lack of good observations of the polar and far-side regions. Observations from a polar vantage point would revolutionize our understanding of the mechanism of solar activity cycles, polar magnetic field reversals, the internal structure and dynamics of the Sun and its atmosphere. Only with extended (many day) observations of the polar regions can the polar flows be determined down to the tachocline where the dynamo is thought to originate. Rapid short period polar orbits, using *in situ* and remote sensing instrumentation, distributed over a small number of spacecraft, will provide continuous 360º coverage of the solar surface and atmosphere in *both* longitude and latitude for years on end. This unprecedented full coverage will enable breakthrough studies of the physical connection between the solar interior, the solar atmosphere, the solar wind, solar energetic particles and the inner heliosphere at large. A potential implementation, the Solar Polar Diamond Explorer (SPDEx) built upon the Solar Polar Imager mission design, involves up to four small spacecraft in a 0.48-AU orbit with an inclination of 75º. The orbit is achieved using solar sails or ion engines, both technologies already demonstrated in space.

**Why Polar Observations?**

As our understanding of the Sun has been revolutionized by a multitude of space missions (Yohkoh, ACE, Ulysses, SOHO, Hinode, STEREO and SDO), the need for information from the polar perspective increases. We do not know the balance between open and closed flux in the heliosphere. We cannot measure the amount of magnetic flux accumulating in the polar regions which is the main precursor of the next solar cycle. We do not fully understand the fine structure of the fast solar wind and the interplay between fast and slow wind in the heliosphere. We do not have a way to predict where new flux will erupt through the solar surface. We have no way of understanding the origin and long-term evolution of solar activity without quantitative information on the polar magnetic flux, its structure and evolution over a significant part of the solar cycle. Moreover, a polar orbit naturally offers observations of the far side, thereby achieving the science goals envisioned for future solar mission concepts in the Solar Physics Decadal Survey (e.g. L5, Safari, etc.) at no additional cost.

**Science Objectives**

Past concept studies have produced a refined set of science objectives that can only be achieved froma short period, long duration, highly inclined polar orbits. We have updated these objectives to align them better with the time horizon for the Next Generation Solar Physics Mission (NGSPM) and relevant missions to be launched soon:

1. *What is the relationship between the magnetism and dynamics of the Sun's polar regions and the solar cycle? More specifically, what is the mechanism of the polar magnetic field reversals, and why and how does the polar field determine the strength of the future solar activity cycle?*
   ESA's Solar Orbiter (SO) mission has a similar objective and is scheduled for a late 2018 launch. However, SO reaches only 34° above the ecliptic and collects imaging observations for only 30 days out of the ~168-day orbits. These orbital constraints enable only a limited time series of polar magnetic field observations that cannot fully address these science objectives. SO will be a valuable trailblazer but will not provide closure.

2. *What is the 3D global structure of the solar corona and how is this influenced by solar activity and coronal mass ejections?*
   The STEREO imagers in combination with SDO have demonstrated the power of multi-viewpoint imaging and 3D reconstructions in revealing non-radial and rotational motions in CMEs, the non-potentiality of active region loops, and the existence of hot magnetic flux ropes within pre-eruptive coronal configurations. But the accuracy of the reconstructions is limited by the ecliptic location of all these imagers. To break the symmetry and avoid the significant line-of-sight effects (particularly around solar minima), observations at latitudes > 45° are needed.

3. *How are variations in the solar wind linked to the Sun at all latitudes? How are solar energetic particles accelerated and transported in radius and latitude?*
   Again, SO will attack these objectives and again its closure is hampered by the mission design. SO reaches maximum latitudes of its highly elliptical orbit at aphelia (~0.7 AU), thus acquiring solar wind measurements over a large range of heliocentric distances. These issues complicate the separation of source and propagation effects in the in-situ solar wind observations. The circular 0.5 AU orbit of our concept will avoid these complications and increase our understanding the solar wind structure in the inner heliosphere.

4. *How does the total solar irradiance vary with latitude?*
   Measurements of the variation of the total solar irradiance (TSI) is fundamental to our understanding both the Sun-Earth Connection and the Sun as a star. The rapid latitude scan of the SPI orbit enables a complete 360º sweep every 4 months with polar and equatorial observations separated by only ~30 days. With parallel modelling efforts, polar brightness contributions and those caused by magnetic activity could be better characterized.

5. *How do polar orbits contribute to Space Weather research?*
   Polar orbits offer a unique advantage for Space Weather (SpWx) research. They combine the benefits of the off Sun-Earth line viewpoints, such as the Lagrangian L5, with a 'bird's-eye' view of the ecliptic plane. Thus, a polar Coronagraph and/or heliospheric imager can follow the changes in the CME speed not only to Earth but also to the inner planets, including Mars. The SPDEx concept provides full coverage of the inner heliosphere all of the time. This capability can become the cornerstone of a robust SpWx system for future robotic and manned exploration.

**Observables**

Unique remote sensing and *in situ* observations made possible by this orbit include:
- Measurements of the time-varying flows, differential rotation and meridional circulation in the polar regions of the Sun down to the tachocline
- Measurements of the polar magnetic field and its temporal evolution
- Monitoring of Earth- and inner planet-directed coronal mass ejections from high latitudes
- Observations of active regions over a significant fraction of their lifetimes
- Measurements of the variation in the total solar irradiance with latitude
- Measurements of chromospheric and low corona outflow velocities vs. structure and latitude
- Measurements of the variation in the magnetic fields, solar wind and solar energetic particles (SEPs) with latitude at constant distance from the Sun

Observing the polar regions of the Sun with a combination of a Doppler-magnetograph and coronal imagers yields opportunities for major new science. Local helioseismology measurements of polar supergranulation flows, differential rotation and meridional circulation, and magnetograms would allow us to understand the mechanisms of the polar field reversals and also the factors determining the amount of the magnetic flux accumulating in the polar regions, which is a primary precursor of the future sunspot cycle. Correlation between measurements of the Doppler signals in the polar regions with disk center measurements from the ground or from near-Earth spacecraft such as SDO, should enable the determination of flows deep within the Sun. When Doppler and magnetograph observations are coupled to UV spectroscopic observations and *in-situ* particle and field measurements, the conceptual SPDEx mission would fundamentally enhance our knowledge of the root causes of solar variability.

**Relevance**

The mission objectives address all aspects of the first challenge and key questions of the 2012 NRC Decadal Report on Solar and Space Physics: *"Understand the structure and dynamics of the Sun's interior, the generation of solar magnetic fields, the origin of the solar cycle, the causes of solar activity, and the structure and dynamics of the corona."* It contributes to the challenge: *"Developing a near-real time predictive capability for understanding and quantifying the impact on human activities of dynamical processes at the Sun".* The mission addresses the 2014 NASA Science Plan Heliophysics goals.

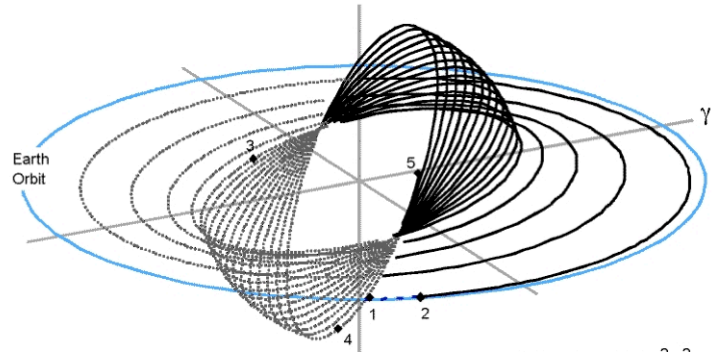

*Figure 1 Conceptual SPDEx polar orbit. The science orbit at 0.48 AU is in a 3:1 resonance with the Earth at an inclination of 75°.*

**Mission Concept & Strawman Payload**

We consider an enhancement of the Solar Polar Imager (SPI) mission concept as described in Liewer et al. (2009) and in the Solar Physics Decadal Survey. The study baselined a solar sail to place a single spacecraft in a 0.48 AU circular orbit around the Sun with an inclination of 75º (Figure 1). A very similar mission concept, POLARIS, was submitted to the ESA Cosmic Vision call (Appourchaux et al., 2009).

**Table 1. Notional SPDEx Payload**

| Remote Sensing | FOV |
|---|---|
| **Doppler Magnetograph** | Full disk |
| **Coronagraph** | 2.5 – 20 Rs |
| **EUV Imager** | 0 – 2 Rs |
| **Heliospheric Imager** | 15 Rs – 1AU |
| **Total Solar Irradiance Monitor** | 2.5º |
| **UV Spectrograph** | 10" x 1.4º |
| **In-situ** | |
| magnetometer | N/A |
| **Composition** | C – Fe, $^{3}$He, $^{4}$He |
| **Energetic particles** | 0.02 – 100 MeV/nuc |

SPDEx uses multiple miniature spacecraft to achieve the same goals. The four spacecraft are arranged in a diamond pattern, 90º apart to provide full 360º coverage. Apart from the magnetographs, the spacecraft do not need to carry identical payloads. For example, only two spacecraft 180º apart need to carry an EUV imager for full coverage. If a coronagraph operates in another ecliptic mission, then a single coronagraph would suffice, etc. To meet the science goals outlined above, we suggest the strawman payload in Table 1, largely drawn from the SPI mission study.

In order to determine the polar flows down to the tachocline, long (many days) nearly continuous observations of the polar regions are required. ***This is the driving requirement for defining a mission with an orbit inclination of 75°.*** The conceptual SPDEx mission could use either a solar sail or an ion engine to reach this nearly polar solar orbit. The final radius is chosen to be 0.48 AU because (a) its orbital period is exactly 1/3 of Earth's (3:1 resonance) allowing an unobstructed Earth-spacecraft line for telecommunications.